\documentstyle[preprint,revtex]{aps}
\begin{document}
\preprint{UM-P-92/113}
\preprint{OZ-92/36}
\preprint{ 1992}
\begin{title}
CP violation in $J/\psi$ $\rightarrow$ $\Lambda$ $\bar {\Lambda}$
\end{title}
\author{Xiao-Gang He, J. P. Ma and Bruce McKellar }
\begin{instit}
Research Center for High Energy Physics\\
School of Physics\\
University of Melbourne \\
Parkville, Vic. 3052 Australia
\end{instit}
\begin{abstract}
We study CP violation in $J/\psi \rightarrow \Lambda \bar{\Lambda}$ decay.
This decay provides a good place to look for CP violation. Some observables
are very sensitive to the $\Lambda$
electric dipole moment $d_\Lambda$ and therefore can be used to improve
the experimental upper bound on $d_\Lambda$. CP violations
in the lepton pair decays of $J/\psi$ and $\Upsilon$ are also discussed.
\end{abstract}
\newpage
Up to now CP violation has only been observed in the neutral Kaon
system\cite{1}.  In
order to isolate the source (or sources) responsible for CP violation,
it is important to find CP violation in other
systems. The measurement of the electric dipole moment of elementary particles
is  very promising as a place to look for further evidence of CP violation.
Stringent experimental upper bound has been
obtained for the neutron\cite{2} and the electron\cite{3}. But the bounds on
$\Lambda$, $\Sigma$ and
other particles are fairly weak.
In this paper we study CP violation in $J/\psi \rightarrow \Lambda
\bar{\Lambda}$ decay. We show that this decay is also a good place to look
for CP violation. Moreover, it can be used to improve the experimental upper
bound on the electric dipole moment of $\Lambda$. We also discuss CP violation
in the lepton pair decays of $J/\psi$ and $\Upsilon$.

The most general decay amplitude A( $J/\psi \rightarrow \Lambda (p_1)
\bar \Lambda (p_2)$) can be parametrized as
\begin{eqnarray}
A(J/\psi \rightarrow \Lambda \bar  \Lambda)
&=& \varepsilon^{\mu}\bar u_{\Lambda}(p_1)[\gamma_{\mu}(a + b\gamma_5) +
(p_{1\mu} -p_{2\mu})(c + id\gamma_5)]v_{\bar {\Lambda}} (p_2)\; ,
\end{eqnarray}
where $\varepsilon^\mu$ is the polarization of the $J/\psi$ particle and in its
rest frame $\varepsilon_\mu = (0, \vec {\bf \varepsilon})$. If CP is conserved,
$d = 0$. The constants
$a,\; b,\;c$ and $d$ are in general complex numbers when
contributions of the absorptive part in the decay amplitude are included.

CP invariance can be tested in $J/\psi \rightarrow \Lambda \bar \Lambda$ only
if
the polarizations of $\Lambda$ and $\bar \Lambda$ can be measured.
Therefore, we will study $J/\psi$ decay to polarized $\Lambda$ ($\bar
\Lambda$). The density matrix for this decay
in the rest frame of $J/\psi$ can be defined as
\begin{eqnarray}
R_{ij} &=&
[\bar u_{\Lambda}(p_1,{\bf s}_1)[\gamma_{i}(a + b\gamma_5) +
(p_{1i} -p_{2i})(c + id\gamma_5)]v_{\bar {\Lambda}} (p_2,{\bf s}_2)\nonumber\\
&\times& \bar v_{\bar \Lambda}(p_2,{\bf s}_2)[\gamma_{j}
(a^* + b^*\gamma_5) +
(p_{1j} -p_{2j})(c^* + id^*\gamma_5)]u_{\Lambda} (p_1,{\bf s}_1)]\; ,
\end{eqnarray}
where $i$ and $j$ label three-vector components.

The CP violating part of this density matrix is
given by
\begin{eqnarray}
R_{ij}&= &r_{ij} + r_{ji}^*\;,\nonumber\\
r_{ij}&= &i2ad^* p_j\{{M^2\over 2} ( {\bf s}_1- {\bf s}_2)_i
- {2M\over {M+2m}} ({\bf s}_1-{\bf s}_2)\cdot{\bf p}  p_i\nonumber\\
&+&imM({\bf s}_1\times {\bf s}_2)_i + i{2M\over {M+2m}}
({\bf s}_1\cdot{\bf p} ({\bf p}\times {\bf s}_2)_i - {\bf s}_2\cdot{\bf p}
({\bf p}\times {\bf s}_1)_i)\}\nonumber\\
&+&2ibd^*M p_j\{ s_{2i}{\bf s}_1\cdot{\bf p} -  s_{1i}
{\bf s}_2\cdot{\bf p}
+i ({\bf p}\times ({\bf s}_1 -{\bf s}_2))_j\}\\
&+&4icd^*M p_i p_j \{-({\bf s}_1 - {\bf s}_2)\cdot {\bf p}
+i({\bf s}_1\times {\bf s}_2)\cdot{\bf p}\}\nonumber\;,
\end{eqnarray}
where ${\bf p}$ is the three momentum of $\Lambda$,
${\bf s}_1$ (${\bf s}_2$) are the polarization vectors of $\Lambda$
($\bar \Lambda$) defined in their rest frames, and $M$ and $m$ are
the masses of $J/\psi$
and the Lambda particles, respectively.
If $J/\psi$ is produced at the threshold at $\bar p\;p$ or $e^+\;e^-$
colliders,
the density matrix $\rho_{ij}$ for the production of $J/\psi$
can be written as
\begin{eqnarray}
\rho_{ij} = {1\over 3} \delta_{ij} + {1\over {2i}} \epsilon_{ijk}
\hat k_kC
-( \hat k_i  \hat k_j - {1\over 3}\delta_{ij})D\;,
\end{eqnarray}
where ${\bf \hat k}$ is the direction of the $p$ or $e$ beam, and $C$ and $D$
are constants which depending on the details of the beams.

In the experimental situation, the polarizations of the Lambda particles
are measured by analysing their decays. We will use
the main decay channels $\Lambda ({\bf s}_1) \rightarrow p({\bf q}_1)
+ \pi^-$ and
$\bar\Lambda ({\bf s}_2) \rightarrow \bar p ({\bf q}_2) + \pi^+$ to
 analyse the polarizations of the Lambda particles.
The density matrix for these two decays in the rest frame of $\Lambda$ ($\bar
\Lambda$), can be written as

\begin{eqnarray}
\rho_\Lambda &= 1 + \alpha_- {\bf s}_1 \cdot {\bf \hat q}_1\;\;\;& \mbox{for}\;
 \Lambda\; \mbox{decay}\;,\nonumber\\
\rho_{\bar \Lambda} &= 1 - \alpha_+ {\bf s}_2 \cdot {\bf \hat q}_2\;\;\;&
\mbox{for}\; \bar\Lambda\; \mbox{decay}\;,
\end{eqnarray}
where $\alpha_- \approx \alpha_+ =0.642\pm 0.013$\cite{4} and ${\bf \hat q}_i =
{\bf q}_i /|{\bf q}_i|$.

Any experimental observables $O$ can be constructed from ${\bf k}$, ${\bf p}$
and ${\bf q}_i$. The expectation value of $O$ is given by
\begin{eqnarray}
<O> &=&{1\over N} {\beta \over {8\pi}}
{1\over {(4\pi)^3}}\int d\Omega_p\;d\Omega_{q_1}\;d\Omega_{q_2}
O \rm {Tr}\{R_{ij}\rho_{ji}\rho_\Lambda \rho_{\bar
\Lambda}\}\;,
\end{eqnarray}
\begin{eqnarray}
N&=& {\beta \over 12\pi}(2|a|^2(M^2+2m^2)+ 2|b|^2(M^2-4m^2)\nonumber\\
&+&|c|^2(M^2-4m^2)^2 +4Re(ac^*)m(M^2-4m^2)\nonumber\\
&=&2M\Gamma (J/\psi \rightarrow \Lambda \bar\Lambda)\;,
\end{eqnarray}
where $\beta = \sqrt{1 - 4m^2/M^2}$, $d\Omega_i$ are the solid angles,
and the trace is over the spins of the Lambda particles.
We find two observables which are particularly interesting,
\begin{eqnarray}
A&=& \theta ({\bf \hat p}\cdot ({\bf \hat q}_1\times {\bf \hat q}_2))
- \theta (-{\bf \hat p}\cdot
({\bf \hat q}_1\times {\bf \hat q}_2))\;,\nonumber\\
B &=& {\bf \hat p}\cdot({\bf \hat q}_1\times {\bf \hat q}_2)\;,
\end{eqnarray}
where $\theta (x)$ is 1 if $x>0$ and is zero if $x<0$. A and B are CP odd
and CPT even observables. Non zero expection
values for them signal CP violation. In terms of the
parameters $a\;, b\;, c\;$ and $d$, we have
\begin{eqnarray}
<A> &=& -{\alpha_-^2 \beta^2 \over 48N}
M^2(2mRe(da^*)+(M^2-4m^2)Re(dc^*))\nonumber\\
<B> &=&-{48\over 27\pi}<A>\;.
\end{eqnarray}
The observable $<A>$ is equal to
\begin{equation}
<A> = {{N^+ -N^-} \over {N^+ +N^-}}\;,
\end{equation}
where $N^{\pm}$ indicate events with $sgn({\bf p}\cdot({\bf q}_1\times
{\bf q}_2)) = \pm$, respectively. $<B>$ measures the correlations between
the momenta.
It is interesting to note that $<A>$ and $<B>$ are insensitive to CP
violation in the Lambda decay amplitude and are independent of the
parameters $C$ and $D$ , e.g. independent of how $J/\psi$ is polarized.
One can also construct CP odd and CPT odd observables from ${\bf p}$, ${\bf k}$
and ${\bf q}_i$. These observables receive contributions from terms
proportional to $Im(da^*)$, $Im(db^*)$ and $Im(dc^*)$ and
CP violation in the $\Lambda$
decay matrix element. It is difficult to analyse these observables.
CP violation in $\Lambda$ decays has been studied in experiments with
$\Lambda$ from $J/\psi$ decays\cite{5}. But in this analysis CP violation
from the $J/\psi$ decay
amplitude was neglected. No study has been performed of the observables
$<A>$ and $<B>$. We will study these two observables in the following.

The branching ratio for $J/\psi \rightarrow \Lambda \bar {\Lambda}$ has been
measured to be $1.35\times 10^{-3}$\cite{4}. From this we can obtain
information about
the parameters in the amplitude. The b-term is a P violating amplitude and is
expected to be significantly smaller than the P conserving a- and c-
amplitudes.
We will therefore neglect contribution from b.
The relative strength of the amplitude a and c can be determined by
studying correlations between the polarization of $J/\psi$ and the direction
of $\Lambda$ momentum. Due to large
experimental errors the constants a and c can not be reliably determined at
present.
In our numerical
estimates we will consider two cases where the decay ampiltude is dominated by
1) the a-term, and 2) the c-term, respectively

The CP violating d-term can receive contributions
from different sources, the electric dipole moment, the CP violating Z
-$\Lambda$ coupling,  etc. In the following we estimate the
contribution from the electric dipole moment $d_{\Lambda}$ of $\Lambda$.
Here $d_{\Lambda}$ is defined by
\begin{equation}
L_{dipole} = i{d_{\Lambda} \over 2} \bar \Lambda \sigma_{\mu \nu}\gamma_5
\Lambda F^{\mu \nu}\;,
\end{equation}
where $F^{\mu \nu}$ is the field strength of the electromagnetic field.
Exchanging a photon between $\Lambda$ and a c-quark, we have the CP violating
c-$\Lambda$ interaction
\begin{equation}
L_{c-\Lambda} = -{2\over 3M^2} e d_{\Lambda}(p_1^\mu - p_2^\mu)\bar c
\gamma_\mu c \bar \Lambda i\gamma_5\Lambda\;.
\end{equation}
{}From this we obtain
\begin{equation}
d =- {2\over 3}  {g_V\over M^2}e d_{\Lambda}\;.
\end{equation}
Here we have used the parametrization,
$<0|\bar c \gamma_\mu c|J/\psi> = \varepsilon_\mu g_V$. The value $|g_V|$ is
determined to be 1.25 GeV$^2$ from $J/\psi \rightarrow \mu^+ \mu^-$.

Inserting the above numbers into equation (9), we obtain
\begin{eqnarray}
|<A>| = \left \{ \begin{array}{ll}
5.6\times 10^{-3}d_\Lambda/  (10^{-16}\;\rm{ecm})\;, &\mbox{if the a-term
dominates}\;\\
1.25\times 10^{-2}d_\Lambda /(10^{-16}\;\rm{ecm})\;, &\mbox{if the c-term
dominates}\;.
\end{array}
\right.
\end{eqnarray}
Here we have used the absolute values for $<A>$ because we can not determine
the relative signs for $a$, $c$ and $d$ from the  experimental data.

The experimental upper bound on $d_{\Lambda}$ is $1.5 \times 10^{-16}
\rm{ecm}$\cite{6}. There are constraints on the strange quark electric
dipole moment and colour dipole moment from the neutron electric dipole
moment $d_n$ \cite{7}, which follow if one assumes that the contributions to
$d_n$ do not cancel against each other. It is
possible that cancellations do occur for $d_n$ but not $d_\Lambda$ and the
constraints from $d_n$ do not necessarily lead to strong constraints on
$d_\Lambda$. Alternative experimental approaches to $d_\Lambda$, such as that
presented here, should therefore be pursued.
If $d_\Lambda$ indeed has a value  close to its experimental upper bound,
the asymmetry $|<A>|$ can be as large as $10^{-2}$.
Of course $<A>$ can also be used to improve bound on $d_\Lambda$. With $10^{7}$
$J/\psi$, it is already possible to obtain some interesting results. This
experiment can be performed with the Beijing $e^+$ $e^-$ machine.
If $10^{9}\; J/\psi$ can be produced, one can improve the  upper bound on
$d_\Lambda$ by an order of magnitude.
This can be achieved in future $J/\psi$ factories.
$|<B>|$ will also give the same information. The same analysis can be easily
applied to $J/\psi$ decays into $\Sigma$, $\Xi$ and etc.

Our analysis can also be used for $J/\psi$ and $\Upsilon \rightarrow
l^+l^-$.
Assuming that the d-term in equation (1) is mainly due to
the electric dipole moment $d_l$ of the leptons, we have
\begin{eqnarray}
<A'> &=& {N'^+ - N'^- \over {N'^+ +  N'^-}}\nonumber\\
&=& {d_l \over e}{\pi \over 4} m_l{\sqrt{1 - 4m_l^2/M^2}\over {1+2m_l/M}}\;,
\end{eqnarray}
where $m_l$ is the lepton mass, and $N'^{\pm}$ is the events with
sgn(${\bf p}\cdot ({\bf s}_1\times {\bf s}_2)$) = $\pm$, respectively.
Here ${\bf s}_i$
are the polarizations of the leptons. For $J/\psi
\rightarrow \mu^+ \mu^-$, we have, $|<A'>| = 4\times 10^{-7}(d_\mu/
{10^{-19}\;\rm{ecm}})$ which is too small to be measured experimentally.
For $\Upsilon \rightarrow
\tau^+ \tau^-$, $|<A'>| = 7\times 10^{-3} d_\tau / (10^{-16}\; \rm{ecm})$.
The experimetal upper bound on $d_\tau$ is $1.6\times 10^{-16}\; \rm{ecm}$
\cite{8},
 so the asymmetry $<A'>$ can be as large as $3\times 10^{-3}$.
Values of $d_\tau$ as large as $10^{-16} \rm{ecm}$ can be obtained in model
calculations. The leptoquark model is one of them\cite{9}. In this model
there is
a scalar which can couple to leptons and quarks. The couplings of the
leptoquark scalar
to the third generation are weakly constrained. It is possible to generate
a large $d_{\tau}$ by exchanging a leptoquark at the one loop level.

\acknowledgments
This work is supported in part by the Australian Research Council.

\end{document}